\documentstyle[preprint,aps,epsf]{revtex}
\begin{document}
\title{Anomalous Mesoscopic Fluctuations
of Transport Coefficients Above The Critical
Temperature }
\author{Fei Zhou$^{a}$, Cristiano Biagini$^{b}$}
\address{$^{a}$Physics Department, Princeton University, Princeton, NJ 08540}
\address{$^{b}$INFM, Unita'di Napoli, Mostra d'Oltremare, 
Pad.19, 80125, Napoli, Italia}
\maketitle

\begin{abstract}
We show in this paper that
above the critical temperature 
of superconductor-metal phase transitions, both the longitudinal 
and Hall conductivity exhibit strong temperature 
dependent mesoscopic fluctuations, with amplitudes 
much larger than the mesoscopic 
fluctuations in noninteracting normal metals.
Such an enhancement of the mesoscopic fluctuations
arises from pairing correlations
and is strongly dependent on dimensions. 
\end{abstract}

\pacs{PACS index numbers: 74.40. +k, 73.23. Ps, 71.30. +h} 

\newpage
\narrowtext
It is well known that the conductance of 
a normal metal exhibits mesoscopic fluctuations
if the sample size $L$ is smaller than the dephasing 
length $L_\phi$$^{[1,2,3]}$. At zero temperature, the amplitude of the 
mesoscopic fluctuations is of order $e^2/\hbar$,
independent of the dimensionality of the sample.
These mesoscopic fluctuations originate from quantum 
interference of electrons and are sensitive
to changes in external magnetic fields,   
impurity configurations, or gate voltages. 

Universal conductance fluctuations({\em UCF}),
of order $e^2/\hbar$,
are closely connected with the universality
of Wigner-Dyson statistics of {\em single} electron levels
in disordered metals.
For a normal metal, the conductance is equal to
$e^2/\hbar$ times $N$, the number of {\em single} electron
levels inside an energy band of the width of the Thouless energy
$E_T$ centered at the Fermi surface. The Thouless energy is 
the inverse of the time required for an electron to diffuse across
the sample$^{[4]}$. 
While the average number of levels within such an energy band
depends on the dimensionality,
$\delta N$,  the 
fluctuation of the number of {\em single}
electron levels within such a band is universally of order of 1$^{[5]}$.
This leads to {\em UCF}. 

At finite temperature, the transport currents are carried by
the quasi particle excitations of energy of order $kT$.
While the total number of electron levels involved is
$N T/E_T$, the amplitude of the fluctuation of the number of 
levels is $(L /L_T)^{d/2}=(T/E_T)^{d/4}$, due to the fact that
the mesoscopic fluctuations of the density of states are
correlated at a length scale $L_T \ll L$ and the 
contributions from different blocks should be summed up
randomly$^{[5]}$. 
Here $L_T=\sqrt{D/T}$
is the normal metal coherence length at temperature $T$, 
$D$ is the diffusion constant. 
The relative amplitude of the fluctuation of number 
of levels decreases as the temperature
is increased.
Therefore the amplitude of the conductance fluctuation
is smaller than $e^2/\hbar$ when the temperature is higher
than $E_T$$^{[5]}$.

The above statement about mesoscopic 
fluctuations of the conductance remains 
true in weakly correlated electron systems. 
For instance, the electron-electron
interaction in normal metals barely 
affect universal conductance fluctuations.
For strongly correlated systems, could be 
fractional quantum Hall systems, or quantum dots in 
the Coulomb blockade regime, the amplitude of the conductance
fluctuation is also of the order   
of or less than $e^2/\hbar$.   

In this paper we study the effect of pairing correlations 
on the mesoscopic fluctuations of the conductance. 
We show that
above the critical temperature, in the presence of pairing 
correlations,  mesoscopic fluctuations of conductance
can greatly exceed $e^2/\hbar$, that of UCF. 
Such an effect increases when the critical 
temperature is approached. It also strongly
depends on dimensionalities of samples,
originating from the fact that pairing correlations 
due to thermal fluctuations strongly depend on  
dimensionalities.

The qualitative mechanism for this phenomenon is as follows.
Above $T_c$ the critical temperature of superconductor-metal
phase transitions,  
there is a finite amplitude for electrons  to 
form superconducting pairs with a certain relaxation time.
Thus transport coefficients in normal metals can be written as 
a sum of classical Drude conductivities and 
contributions arising from pairing correlations.
The amplitude of thermal 
fluctuations of superconducting pairs 
are determined by a competition between
the entropy and the condensation energy
and becomes divergent when  
the temperature approaches $T_c$ from above. 
The typical relaxation time is given 
as the time scale for pairs to diffuse 
over the Landau-Ginsburg length scale and
is also divergent when the critical temperature is 
approached. 
As a result, the conductivity is enhanced via 

\begin{eqnarray}
&&\frac{\delta \sigma}{\sigma} \propto 
\int \frac{d{\bf Q}^d}{(2\pi)^d} \Delta_{\bf Q} \Delta_{\bf Q} 
\tau_{\bf Q}
= \frac{1}{g_d}\left(\frac{T}{T-T_c}\right)^{(4-d)/2} 
\nonumber \\
&&\Delta_{\bf Q} \Delta_{\bf Q}=
\nu^{-1}\left(\frac{T-T_c}{T} + \frac{D{\bf Q}^2}{T}\right)^{-1},
\tau_{\bf Q}=\frac{1}{D{\bf Q}^2 + T-T_c}. 
\end{eqnarray} 
where $\nu$ is the density of states in normal metals;  
$g_d$ is the dimensionless conductance
of the size of $L_T$, and $d$ is the dimensionality.  
The dimensionless conductance in each dimension is
$g_3=2\pi^2\hbar\sigma L_T/e^2$, $g_2=4\pi\hbar \sigma t/e^2$ and 
$g_1=6\pi\hbar \sigma a^2/L_T e^2$,
$t$ is the film thickness in $2D$, $a$ is the diameter in 
$1D$.  $\sigma=e^2\nu D$ is the Drude conductivity. Integral 
$\int d{\bf Q}^d/(2\pi)^d$ is $\int d{\bf Q}^3/(2\pi)^3$ in $3D$,
$1/t \int d{\bf Q}^2/(2\pi)^2 $ in $2D$ and $1/a^2 \int d{\bf Q}/(2\pi)$
in $1D$.  Eq.1 is valid as far as the correction is
small .i.e., $\delta \sigma/\sigma \ll 1$. 

However, the condensation energy 
has mesoscopic fluctuations, as emphasised in Ref. 6.
The fluctuation amplitude is $1/g_d$ and its correlation length
is min$\{ L_T, \xi_0\}$. Here $\xi_0=\sqrt{\hbar D/T_c}$ 
is the coherence length of superconductors at zero temperature.
This effectively leads to  
mesoscopic fluctuations of the critical temperature,
${\delta T_c}/{T_c} \propto g_d^{-1}({\xi_0}/{L})^{4-d/2}$
where we take into account that $L \gg \xi_0$ and
fluctuations from different blocks of the size
of $\xi_0$ should be summed up randomly. 
Therefore the pairing amplitude calculated in Eq.1 
develops giant mesoscopic fluctuations $\delta \Delta^M_{\bf Q}
\delta\Delta^M_{\bf Q}$ near the critical point,

\begin{equation}
\frac{\delta \Delta^M_{\bf Q} 
\delta \Delta^M_{\bf Q}}{\Delta_{\bf Q} \Delta_{\bf Q}} 
\propto 
\left(\frac{1}{g_d}\right) \left(\frac{\xi_0}{L}\right)^{4-d/2}
\left(\frac{T}{T- T_c +D{\bf Q}^2}\right).
\end{equation}
Substituting Eq.2 into $\delta \sigma$  in Eq. 1 
we obtain an estimate of  
mesoscopic fluctuations of the conductance, 
of the same order as that given in Eq.8.
Mesoscopic fluctuations of the conductance in this regime are 
therefore determined by mesoscopic fluctuations of
pairing correlations. 
This results in anomalous mesoscopic fluctuations
of transport coefficients in such systems.

To proceed further, we express
the conductivity of a given sample above $T_c$ 
as $\sigma_{xx}=\sigma +\delta\sigma+\delta \sigma^M_{xx}$.
$\delta\sigma$ represents the contributions from Aslamazov-Larkin
and Maki Thompson corrections due to thermal fluctuations,
studied in Refs. 7.8. It is divergent as the temperature
approaches $T_c$, as shown in Eq.1. 
The mesoscopic fluctuations of the conductivity $\delta\sigma^M_{xx}$ 
as a function of gate voltage $V_g$ are given in terms of diagrams in Fig.a),

\begin{eqnarray}
&&<\delta \sigma^M_{xx}(V_{g1}) 
\delta \sigma^M_{xx}(V_{g2})>=
<\delta\sigma^M_{AL}(V_{g1})
\delta\sigma^M_{AL}(V_{g2})> + 
<\delta\sigma^M_{MT}(V_{g1})
\delta\sigma^M_{MT}(V_{g2})>,
\nonumber \\  
\nonumber \\ 
&&<\delta\sigma^M_{AL}(V_{g1})
\delta\sigma^M_{AL}(V_{g2})>  
=\left(\frac{4 e^2}{\hbar \pi}\right)^2\left(\frac{\pi D}{8 T_c}\right)^4 
\int \frac{d{\bf Q_1}^d}{(2\pi)^d}
\frac{d{\bf Q_2}^d}{(2\pi)^d}
{\bf Q_1}^2 {\bf Q_2}^2 
\int^{+\infty}_{-\infty}
d\omega_1 d\omega_2 
\coth \left(\frac{\omega_1}{2T}\right) 
\coth \left(\frac{\omega_2}{2T}\right)\nonumber \\ 
&&\frac{\partial^2} {\partial \omega_1 \partial \omega_2}
\left\{Im  L^R( \omega_1, {\bf Q_1}^2) 
Im L^R( \omega_2, {\bf Q_2}^2) 
<Im \delta L^R(\omega_1, {\bf Q}_1^2) Im \delta L^R(\omega_2, {\bf Q}_2^2)>
\right\},
\nonumber \\  
\nonumber \\  
&&<\delta\sigma^M_{MT}(V_{g1})
\delta\sigma^M_{MT}(V_{g2})>
=\left(\frac{4 e^2}{\hbar \pi}\right)^2 D^2 
\int \frac{d{\bf Q_1}^d}{(2\pi)^d}
\frac{d{\bf Q_2}^d}{(2\pi)^d}
{\cal C}_0({\bf Q_1}^2)
{\cal C}_0({\bf Q_2}^2)
\int_0^{\infty}d\omega_1 d\omega_2
\coth\left(\frac{\omega_1}{2T}\right)  
\coth\left(\frac{\omega_2}{2T}\right)  
\nonumber \\
&&\frac{\partial^2}{\partial\omega_1 \partial\omega_2}
\left\{ Im\Psi\left(\frac{1}{2} +\frac{i\omega_1}{4\pi T}\right)
Im\Psi\left(\frac{1}{2} +\frac{i\omega_2}{4\pi T}\right)
<Im \delta L^R(\omega_1, {\bf Q_1}^2) 
Im \delta L^R(\omega_2, {\bf Q_2}^2)> 
\right \}.
\end{eqnarray}
Here the lowercase $AL$ and $MT$ represent 
mesoscopic fluctuations of Aslamazov-Larkin
and Maki-Thompson corrections to the conductivity respectively.
We neglect mesoscopic fluctuations of conductivities
associated with normal quasi-particles as the temperature is close to $T_c$ . 
$<...>$ denotes the average over impurity scattering potentials.
$\Psi$ is the digamma function. $Im \delta L^R=(\delta L^R -\delta L^A)/2i$.
The propagators $L^{R,A}$, $\delta L^{R, A}$ are defined as 

\begin{eqnarray}
&&L^{R, A}(\omega, {\bf Q}^2)=\left(\frac{T-T_c}{T} +\frac{\pi}{8}
\frac{D{\bf Q}^2  \pm i\omega}{T}
+\frac{\omega}{2} \frac{\partial \ln T_c}{\partial\epsilon_F}
\right)^{-1}
\nonumber \\
&&\delta L^{R, A}(\omega, {\bf Q}^2)=
L^{R,A}(\omega, {\bf Q}^2)^2
\int d\epsilon \tanh(\frac{\epsilon}{2kT})
\nonumber \\
&& \frac{1}{\nu V}\int d{\bf r} d{\bf r}'
\exp(i{\bf Q} \cdot ({\bf r} -{\bf r}'))
\left \{ G^R_{\epsilon+\omega}({\bf r}, {\bf r}')
G^A_{-\epsilon}({\bf r}, {\bf r}') -
<G^R_{\epsilon+\omega}({\bf r}, {\bf r}')
G^A_{-\epsilon}({\bf r}, {\bf r}')> \right\},
\end{eqnarray}
where $V$ is the volumn of the sample.
$G^{R, A}$ are the exact retarded and advanced Green function
in the presence of disorder.   
$\delta L^{R, A}(\omega_1, {\bf Q}_1^2)$,
$\delta L^{R, A}(\omega_2, {\bf Q}_2^2)$
are evaluated in the presence of gate voltages $V_{g1}$ and
$V_{g2}$ respectively. 
$<\delta L^{R, A}>=0$ and the 
correlation function is given as

\begin{eqnarray}
&&{<\delta L^{R,A}(\omega_1, {\bf Q}_1^2) 
\delta L^{R,A}(\omega_2, {\bf Q}_2^2)>}
=L^{R,A}(\omega_1, {\bf Q}_1^2)^2 L^{R,A}(\omega_2, {\bf Q}_2^2)^2
\nonumber \\
&&\int_{-\infty}^{+\infty} d\epsilon d\epsilon'  
\frac{1}{2\epsilon}
\frac{1}{2\epsilon'}
\tanh(\frac{\epsilon}{2T})
\tanh(\frac{\epsilon'}{2T})
\frac{1}{\nu^2 V^2}\int
d{\bf r}d{\bf r}'
\{ Re {\cal C}^2_{\epsilon-\epsilon'}({\bf r}, {\bf r'})
+ Re {\cal D}^2_{\epsilon-\epsilon'}({\bf r}, {\bf r'})\}
\end{eqnarray}
in the leading order of $T-T_c/T_c$.
${\cal D}_{\epsilon-\epsilon'}({\bf r}, {\bf r'})
=\nu^{-1}<G^R_{\epsilon}({\bf r}, {\bf r'})
G^A_{\epsilon'}({\bf r}', {\bf r})>$,
${\cal C}_{\epsilon-\epsilon'}({\bf r}, {\bf r'})
=\nu^{-1}<G^R_{\epsilon}({\bf r}, {\bf r'})
G^A_{\epsilon'}({\bf r}, {\bf r}')>$,
are the diffusons and cooperons. Generally, they satisfy 

\begin{equation}
\{ i\omega +ie\phi_1({\bf r}) -ie\phi_2({\bf r})
+D(i\hbar\nabla -\frac{e}{c}{\bf A}_1({\bf r}) \pm \frac{e}{c}
{\bf A}_2({\bf r}))^2 \}{\cal D}({\cal C})_{\omega}({\bf r}, {\bf r}') 
=\delta({\bf r}- {\bf r'})
\end{equation}
and ${\cal D}={\cal C}=0$ at the boundary for open geometry samples$^{[1,2]}$.
$\phi_{1(2)}$ is the electrical potential induced via
the gate voltage $V_{g1(2)}$ and ${\bf A}_{1(2)}$ is the vector potential 
in the presence of a magnetic field. 
${\cal C}_0({\bf Q}^2)$ in Eq.3 is the Fourier transformation
of ${\cal C}_0({\bf r}, {\bf r}')$.
The last term in the expression
of $L^{R, A}$ in Eq.4 is from the dependence of the
Fermi energy, $\epsilon_F$,
of the critical temperature, $T_c$, and is inversely 
proportional to $\epsilon_F$. It is important only when 
the Hall conductivity is concerned$^{[9, 10]}$. 

At $V_{g1}=V_{g2}$ in the absence of magnetic fields, Eq.5 yields

\begin{equation}
<\delta L^{R,A}(\omega_1, {\bf Q}_1^2) 
\delta L^{R,A}(\omega_2, {\bf Q}_2^2)>
=L^{R,A}(\omega_1, {\bf Q}_1^2)^2 L^{R,A}(\omega_2, {\bf Q}_2^2)^2
\frac{\alpha_d}{g_d^2}
\left(\frac{L_T}{L}\right)^{4-d}.
\end{equation}
Here $\alpha_d$ are the constants dependent on the geometry and 
the dimensionality of the sample.
When $\xi(T)=\sqrt{D/(T-T_c)} \gg L$, in the open geometry
in which we are interested, 
the most divergent contribution 
to the Aslamazov-Larkin and the Maki Thompson corrections
to the conductivity
is determined by the fluctuations with ${\bf Q}=\pi/L$, i.e,
$L^R(\omega, {\bf Q}=\pi/L)$.
Substituting Eq.7 into Eq.3, 
we obtain the amplitude of mesoscopic fluctuations of the
conductivity above the critical temperature

\begin{equation}
\frac{<(\delta\sigma_{xx}^M)^2>}{\sigma^2}=
\frac{\beta_d}{g^4_d}
\left(\frac{T}{T- T_c}\right)^{\frac{8-d}{2}}
\left(\frac{\xi(T)}{L}\right)^{4-d},
\end{equation}
when $L \gg \xi(T)$ and saturates as

\begin{equation}
\frac{<(\delta\sigma_{xx}^M)^2>}{\sigma^2}=
\frac{\beta_d}{g^4_d}
\left(\frac{T}{E_T}\right)^{\frac{8-d}{2}},
\end{equation} 
when $L \ll \xi(T)$.  
Here $\beta_3\propto 1$, $\beta_2 \propto 
max\{\ln(L/\xi(T)), 1\}$
and $\beta_1 \propto max\{L/\xi(T), 1\}$.
We want to emphasize that the mesoscopic 
fluctuations discussed here strongly depend on the dimensionality  
of a sample, which is in contrast to the theory of $UCF$.
This is a direct consequence of mesoscopic fluctuations of
pairing correlations. 

Eqs.8, 9 are valid as far as $\delta \sigma^{M}_{xx} \ll \delta\sigma
\ll \sigma$.
Following Eqs.8,9, mesoscopic fluctuations
of conductances can be much larger than $e^2/\hbar$.
For instance, for a $2D$ film of the size of $\xi(T)$,
at the temperature $T- T_c \sim T_c/g_2$
when $\delta \sigma/\sigma \sim 1$, 

\begin{equation}
\sqrt{<(\delta\sigma_{xx}^M)^2>} \propto 
\frac{e^2}{t\hbar}\sqrt{g_2}
\end{equation}
is parametrically larger than 
UCF in normal metals.

The anomalous fluctuations can be probed in  
experiments where resistances are  measured 
at different gate voltages. Let us consider a $2D$ film 
where a gate voltage is applied to the top of the film 
with capacitance $C$. The
electric field induced by the gate is normal to the film
and is screened over a Debye screening length $r_0=(e^2\nu)^{-1/2}$. 
Substituting Eqs. 5, 6 at ${\bf A}_1={\bf A}_2=0$ into Eq.3, we obtain
the gate voltage dependence of the mesoscopic fluctuations

\begin{eqnarray}
{<(\delta\sigma_{xx}^M(V_{g1})-\delta\sigma^M_{xx}(V_{g2}))^2>}
=<(\delta\sigma_{xx}^M)^2>F(\frac{|V_{g1} -V_{g2}| C r^2_0}{\epsilon_0 t L^2 T})
\nonumber \\
\end{eqnarray}
where
\begin{equation}
F(x)\propto \left\{ \begin{array}{cc}
x, &  \mbox{$ x \ll 1$} \\
1, &  \mbox{$ x \gg 1$} 
\end{array}
\right.
\end{equation}
Following Eq.11, in this case, the characteristic 
gate voltage $V_g$ 
at which $\delta\sigma^M_{xx}(V_g)$ are correlated is 
$\epsilon_0 T L^2 t/e r^2_0 C$.
Mesoscopic fluctuations discussed here are also sensitive to  
external magnetic fields. At  
$E_T={D/L^2} \ll DH \ll T -T_c$, 
we can neglecte the magnetic field dependence of $L^{R, A}$
in the leading order of $DH/(T-T_c)$. 
As a result, the correlation of 
conductance fluctuations as a function of magnetic field 
is determined by Eq.3, with $<Im \delta L^{R}(V_{g1})
Im \delta L^{R}(V_{g2})>$ replaced with
$<Im \delta L^{R}({\bf H})
Im \delta L^{R}(0)>$. Taking into account Eqs.5,6
at $\phi_1=\phi_2$, we obtain 
conductance fluctuations
of a 2D film as functions of a magnetic field ${\bf H}$ perpendicular
to the film

\begin{equation}
\frac{<(\delta\sigma_{xx}^M({\bf H})-\delta\sigma^M_{xx}(0))^2>}
{<\delta\sigma_{xx}^M\delta \sigma_{xx}^M>}
\propto \frac{D H}{T}.
\end{equation}
Eq. 13 is valid when $E_T \ll DH \ll T -T_c$ and saturates when
$DH \gg T-T_c$. 
Eq. 13 shows that $\sigma(H)$ diffuses in $H$ space, 
with diffusion constant $<(\delta\sigma_{xx}^M)^2> L_T^2/
\Phi_0$,
and the mean free time $\Phi_0L^{-2}$,
$\Phi_0=\hbar c/e$ is the flux quantum. 
However, the average pairing correlation 
is also suppressed in the presence of external fields$^{[7,8,9,10]}$,  
with the characteristic magnetic field corresponding
to one flux per area of the size $\xi^2(T)$.
Thus, in conductivity measurements,
the dependence of mesoscopic fluctuations on magnetic fields  
should be differentiated from
the average magnetoresistance. The other possibility 
to observe the anomalous mesoscopic fluctuations
of transport coefficients is to measure the conductance 
during different thermal cycles.

Let us now turn to mesoscopic fluctuations
of the Hall conductivity.
Again one can write for a given sample,
$\sigma_{xy}=\tau\Omega_c \sigma +\delta \sigma_{xy} +\delta\sigma^M_{xy}$.
The first term is the Hall conductivity obtained 
from the classical Boltzmann transport equation,
where $\Omega_c =e H/mc$ is the cyclotron frequency.
The second term is the correction to the classical result due to  
pairing correlations above $T_c$, 
$\delta \sigma_{xy} \propto \tau\Omega_c \sigma/g_d (T/T-T_c)^{(6-d)/2}$ 
as calculated in Ref.9, 10. 
$\delta\sigma^M_{xy}$ represents mesoscopic fluctuations of the Hall 
conductivity.
The propagator in the presence of an 
external magnetic field is determined by diagrams in Fig.b),
\begin{equation}
L^R(\omega, {\bf Q}) 
\frac{D {e {\bf A}_H \cdot {\bf Q}}}{T} 
L^R(\omega, {\bf Q})
\end{equation}
proportional to
${\bf A}_H \cdot {\bf Q}$.
This leads to one more $L^R(\omega, {\bf Q})$ in the expression
for Hall conductivity than   
in Eq.3 and yields a more divergent
temperature dependence of $\delta\sigma^M_{xy}$ as $T_c$ is approached. 
Here ${\bf A}_H$ is the vector potential of the external
magnetic field ${\bf H}={\bf q}\times {\bf A}_H$, ${\bf E}$ is the electrical
field.  Noticing that ${\bf A}_H \cdot {\bf Q}
{\bf E} \cdot {\bf q} {\bf Q}= {\bf Q}^2 {\bf E} \times {\bf H}$,
taking into account the gradient of
$T_c$ at the Fermi surface in the 
fluctuation propagtors, as shown in Eq.4, in the leading order
of $T_c/\epsilon_F$,  
we obtain mesoscopic fluctuations of the Hall conductivity,

\begin{eqnarray}
<(\delta\sigma^M_{xy})^2>= (\tau\Omega_c)^2\frac{e^4}{\hbar^2} 
\left(\frac{\pi D}{8 T^2}\right)^4
\int \frac{d{\bf Q_1}^d}{(2\pi)^d} 
\frac{d{\bf Q_2}^d}{(2\pi)^d} 
{\bf Q_1}^2 {\bf Q_2}^2
\int_{-\infty}^{+\infty}d\omega_1 d\omega_2 
\frac{\omega_1
\omega_2}{\sinh^{2}\left(\frac{\omega_1}{2T}\right)
\sinh^{2}\left(\frac{\omega_2}{2T}\right)} 
\nonumber \\
((Im L^R(\omega_1, {\bf Q}_1^2))^3
(Im L^R(\omega_2, {\bf Q}_2^2))^3
<Re \delta L^R(\omega_1, {\bf Q}_1^2)
Re\delta L^R(\omega_2, {\bf Q}_2^2)>+
9 (Im L^R(\omega_1, {\bf Q}_1^2))^2 
\nonumber \\
\times (Im L^R(\omega_2, {\bf Q}_2^2))^2
Re L^R(\omega_1, {\bf Q}_1^2)
Re L^R(\omega_2, {\bf Q}_2^2)
<Im \delta L^R(\omega_1, {\bf Q}_1^2) 
Im \delta L^R(\omega_2, {\bf Q}_2^2)>).
\end{eqnarray}
We neglect the Maki-Thompson contribution
to the Hall conductivity because it is less divergent 
than the result in Eq.15.

Substituting Eq.5 into Eq.15, we obtain 

\begin{equation}
\frac{<(\delta\sigma_{xy}^M)^2>}{\sigma^2}
=(\tau\Omega_c)^2
\frac{\gamma_d}{g^4_d}
\left(\frac{T}{T- T_c}\right)^{\frac{12-d}{2} }
\left(\frac{\xi(T)}{L}\right)^{4-d}, 
\end{equation}
when $L \gg \xi(T)$; When $L \ll \xi(T)$, 

\begin{equation}
\frac{<(\delta\sigma_{xy}^M)^2>}{\sigma^2} 
=(\tau\Omega_c)^2
\frac{\gamma_d}{g^4_d}
\left(\frac{T}{E_T}\right)^{\frac{12-d}{2}}. 
\end{equation}
Here $\gamma_d$ is a constant of order of unity, dependent on
dimensionalities of samples.
Eqs. 16, 17 are valid when
$\delta \sigma_{xy} \ll \tau\Omega_c \sigma$, i.e.
$T- T_c \gg T_c/g^{2/(6-d)}_d$.
For a $2D$ sample of the size of order of $\xi(T)$,
at the temperature $T- T_c \sim T_c/g^{1/2}_2$,

\begin{equation}
\sqrt{<(\delta\sigma_{xy}^M)^2>}
\propto \tau\Omega_c \frac{e^2}{t\hbar} g_2^{\frac{1}{4}}. 
\end{equation}
Generally speaking, in a disordered mesoscopic 
sample, mesoscopic fluctuations
of the transverse conductivity are nonzero even in the absence
of an external magnetic field$^{[3]}$.
However, as usual, by reversing 
directions of external magnetic fields,
one can measure the Hall conductivity and its  mesoscopic fluctuations
as the asymmetrical part of $\sigma_{xy}({\bf H})$.

In conclusion, we would like to point out that 
in a few recent experiments, 
giant mesoscopic fluctuations of conductance have been found in  
granular superconductors above their critical temperatures$^{[11, 12]}$.
We believe that the mechanism discussed in this paper is relevant
for those phenomena.  At zero temperature,
it was shown that mesoscopic fluctuations
of the condensation energy 
could lead to a superconducting glass state$^{[6]}$.
F. Zhou would like to thank Aviad Frydman for sending him the preprint
of his group's work and Y. Liu for showing him unpublished data. 
We acknowledge illuminating discussions with B. Altshuler, 
B. Spivak, A. A. Varlamov. F. Zhou
is supported by Princeton University.
C. Biagini is supported in part under PRA97-QTMD in Italy.
We also like to thank NEC research Institute for its hospitality.

\begin{figure}

\begin{center}
\leavevmode
\epsfbox{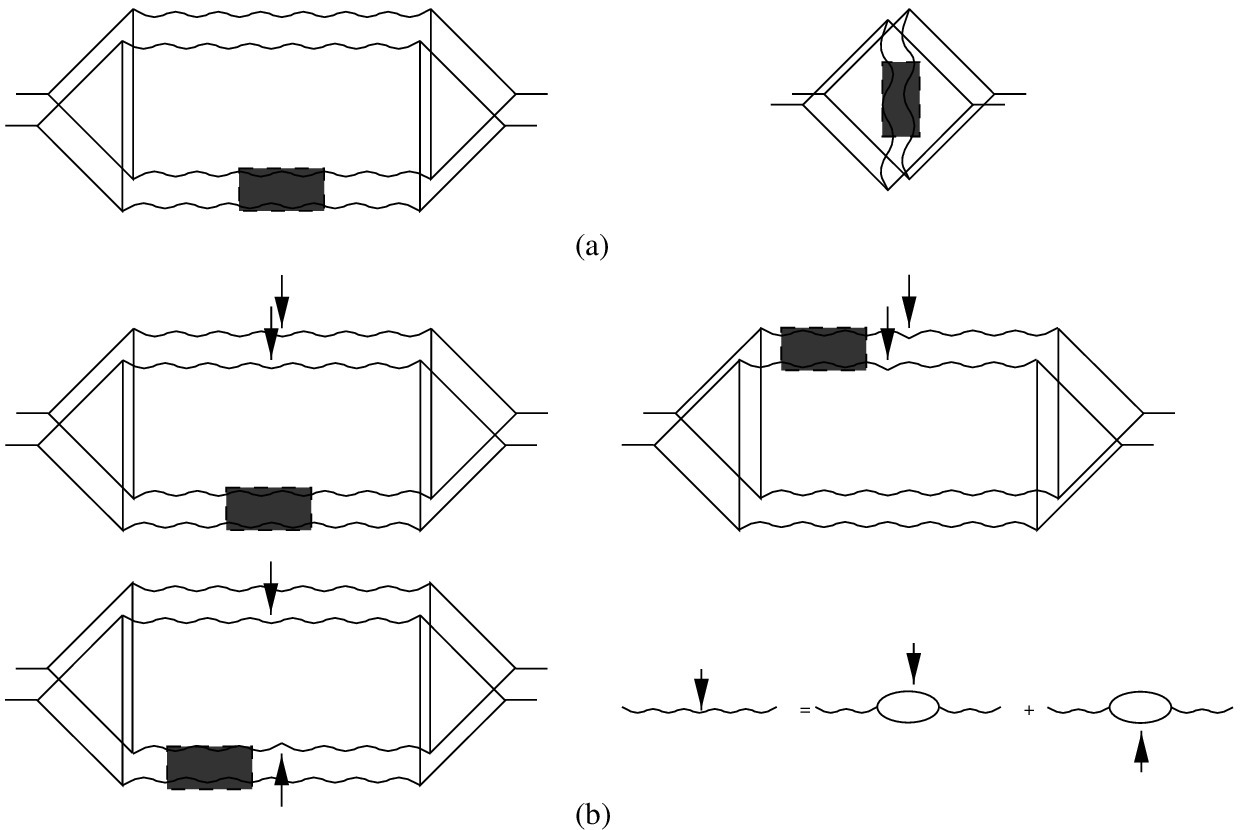}
\end{center}
\caption{Diagrams for anomalous mesoscopic fluctuations
of transport coefficients above the critical temperature.
In a), b), the solid lines represent electron
Green functions, 
the wave lines represent propagator $L^{R, A}$;
the wave lines with
shaded boxes represent $<\delta L^{R, A}\delta L^{R,A}>$, given in Eq.5. 
In b), the wave lines carrying arrows represent 
the propagators in the presence of magnetic fields.
Triangles stand for current vertices and
external electrical field vertices.}
\end{figure}

\end{document}